\documentclass[fleqn,usenatbib]{mnras}

\usepackage{newtxtext,newtxmath}

\usepackage[T1]{fontenc}

\usepackage{bm}

\DeclareRobustCommand{\VAN}[3]{#2}
\let\VANthebibliography\thebibliography
\def\thebibliography{\DeclareRobustCommand{\VAN}[3]{##3}\VANthebibliography}

\usepackage{graphicx}	
\usepackage{amsmath}	

\title[Neutrino emission]{Neutrino emission from FRB-emitting magnetars}

\author[Y. Qu and B. Zhang]{
Yuanhong Qu$^{1}$\thanks{E-mail: yuanhong.qu@unlv.edu}
and
Bing Zhang$^{1}$\thanks{E-mail: bing.zhang@unlv.edu}
\\
$^{1}$Department of Physics and Astronomy, University of Nevada Las Vegas, Las Vegas, NV 89154, USA
}

\date{Accepted XXX. Received YYY; in original form ZZZ}

\pubyear{2021}

\begin{document}
\label{firstpage}
\pagerange{\pageref{firstpage}--\pageref{lastpage}}
\maketitle

\begin{abstract}
The detection of a bright radio burst (hereafter FRB 200428) in association with a hard X-ray burst from the Galactic magnetar SGR 1935+2154 suggests that magnetars can make FRBs. We study possible neutrino emission from FRB-emitting magnetars by developing a general theoretical framework. We consider three different sites for proton acceleration and neutrino emission, i.e. within the magnetosphere, in the current sheet region beyond the light cylinder, and in relativistic shocks far away from the magnetosphere. 
Different cooling processes for protons and pions are considered to calculate the neutrino emission suppression factor within each scenario. We find that the flux of the neutrino emission decreases with increasing radius from the magnetar due to the decrease of the target photon number density. We calculate the neutrino flux from FRB 200428 and its associated X-ray burst. The flux of the most optimistic case invoking magnetospheric proton acceleration is still $\sim4$ orders of magnitude below the IceCube sensitivity. We also estimate the diffuse neutrino background  from all FRB-emitting magnetars in the universe. The total neutrino flux of magnetars during their FRB emission phases is a negligible fraction of observed diffuse emission even under the most optimistic magnetospheric scenario for neutrino emission. However, if one assumes that many more X-ray bursts without FRB associations can also produce neutrinos with similar mechanisms, magnetars can contribute up to $10^{-8} \ {\rm GeV \ s^{-1} \ sr^{-1} \ cm^{-2}}$ diffuse neutrino background flux in the GeV to multi-TeV range. 
\end{abstract}




\section{Introduction}

Fast Radio Bursts (FRBs) are bright radio bursts with extremely high brightness temperatures and millisecond durations \citep{Lorimer07,Thornton13,CHIME19}. The detection of a bright millisecond radio burst (hereafter FRB 200428) with properties similar to those of FRBs from SGR 1935+2154 \citep{Bochenek2020,CHIME/FRB2020} and its association with a hard X-ray burst \citep{CKLi20,Mereghetti20,konus,AGILE} suggested that magnetars \citep{Thompson&Duncan1995,Thompson&Duncan1996} could be responsible for at least some, probably most FRBs observed in the universe \citep{Popov10,Thornton13,Lyubarsky14,Katz16,Metzger17,Metzger19,Kumar17,Beloborodov17,Beloborodov20,YangZhang18,Wadiasingh20,Lu20,Margalit20,Lyubarsky20,Yang20,YangZhang21}. Within the magnetar models for FRBs, two types have been discussed in the literature \citep{Zhang2020nature,Lyubarsky21}: the pulsar-like models invoking coherent emission within or just outside the magnetosphere \citep[e.g.][]{Kumar17,Lu&Kumar18,YangZhang18,Wadiasingh20,Lu20,Lyubarsky20,YangZhang21} and the GRB-like models invoking relativistic shocks far away from the magnetosphere \citep[e.g.][]{Lyubarsky14,Beloborodov17,Metzger19,Beloborodov20,Margalit20,Yu21}.

On the other hand, astrophysical sources of cosmic high-energy neutrino background \citep{IceCube13} has not been identified. Theoretically, many possible sources have been discussed to generate cosmic neutrinos  \citep[e.g.][]{Meszaros17,Murase&Bartos19}. Observationally, blazars \citep{IceCube2018} and tidal disruption events \citep{Stein21} have been suggested as the possible candidates, but the latest analyses show that the significance of these associations is not strong \citep[e.g.][]{Luo&Zhang20PRD,IceCube21}.

The possibility that magnetars could be TeV neutrino emitters has been suggested by \cite{Zhang03} within the framework of magnetospheric photonmeson interaction mechanism. \cite{Murase09} discussed neutrino emission from young magnetars due to $pp$-interaction between accelerated protons and the stellar ejecta. No evidence of neutrino emission from Galactic magnetars has been collected so far \citep{Ghadimi2021}. \cite{Metzger20} calculated the neutrino flux of FRB 200428-like Galactic FRBs within the framework of the external shock model of FRBs and showed that it is many orders of magnitude below the IceCube sensitivity unless the FRB is much brighter and closer to Earth.

In this paper, we reinvestigate the neutrino emission problem for FRB emitting magnetars within a generic framework invoking proton acceleration and photomeson interactions with X-ray photons associated with FRBs. Specifically, we consider relativistic protons in three possible acceleration sites, namely, within the magnetar magnetosphere, in the current sheet region, and in relativistic shocks.

This paper is organized as follows. In section 2, we briefly introduce the physics of photomeson interactions and three possible mechanisms to accelerate protons to high Lorentz factors. In section 3, we perform an estimate of neutrino emission power for the three scenarios. In section 4, we calculate the detailed neutrino emission spectra from FRB 200428 by carefully considering the flux suppression effect due to various cooling processes of protons and pions. In section 5, we present the diffuse neutrino emission from FRB-emitting magnetars across the universe. The main conclusions and discussions are summarized In section 6. The convention $Q=10^nQ_{n}$ in cgs units is adopted. The cosmological parameters $H_0=67.8 \ \rm{km} \ \rm{s^{-1}} \ \rm{Mpc^{-1}}$, $\Omega_{M}=0.308$, and $\Omega_{\Lambda}=0.692$  \citep{Planck16} are applied.

\section{Photomeson interaction and proton acceleration sites}
\label{mechanism}

\subsection{Photomeson interaction threshold condition}
Accelerated protons can interact with photons to produce neutrinos via the Delta-resonance
$p\gamma\rightarrow\Delta^{+}\rightarrow n\pi^{+}\rightarrow n\nu_{\mu}\mu^{+}\rightarrow n\nu_{\mu}e^{+}\nu_e\bar\nu_{\mu}$. The kinematic condition is
\begin{equation}\label{matching}
\epsilon_p\epsilon_{\gamma}\geq 0.32(\rm{GeV})^2(1-\cos\theta_{p\gamma})^{-1},
\end{equation}
where $\theta_{p\gamma}$ is the incident angle between proton and photon.

One can use the observational data of X-rays to estimate the proton energy needed to produce neutrinos. For FRB 200428, the associated X-rays have energy 
$\epsilon_{\gamma} \simeq 10-200 \ \rm{keV}$ \citep{CKLi20}, which requires
\begin{equation}
\epsilon_{p}\geq (3.2 \ {\rm TeV}) \ \epsilon_{\rm \gamma,100 keV}^{-1} f_g,
\end{equation}
where $f_g$ is a geometric factor, $\epsilon_p$ is proton energy, $\epsilon_{\gamma}$ is photon energy, which is normalized to 100 keV.
For $p\gamma$ interactions, about 0.2 times of the proton energy goes to $\pi^+$, whose energy is evenly split to four leptons. So each neutrino would have an energy $\epsilon_\nu \sim 0.05 \epsilon_p$. Since the 10 keV X-rays correspond to $\epsilon_p$ up to $\sim 32 \ {\rm TeV}$, FRB 200428 can produce neutrinos with energy up to at least $\sim 1.6$ TeV.

\begin{figure}
	\includegraphics[width=\columnwidth]{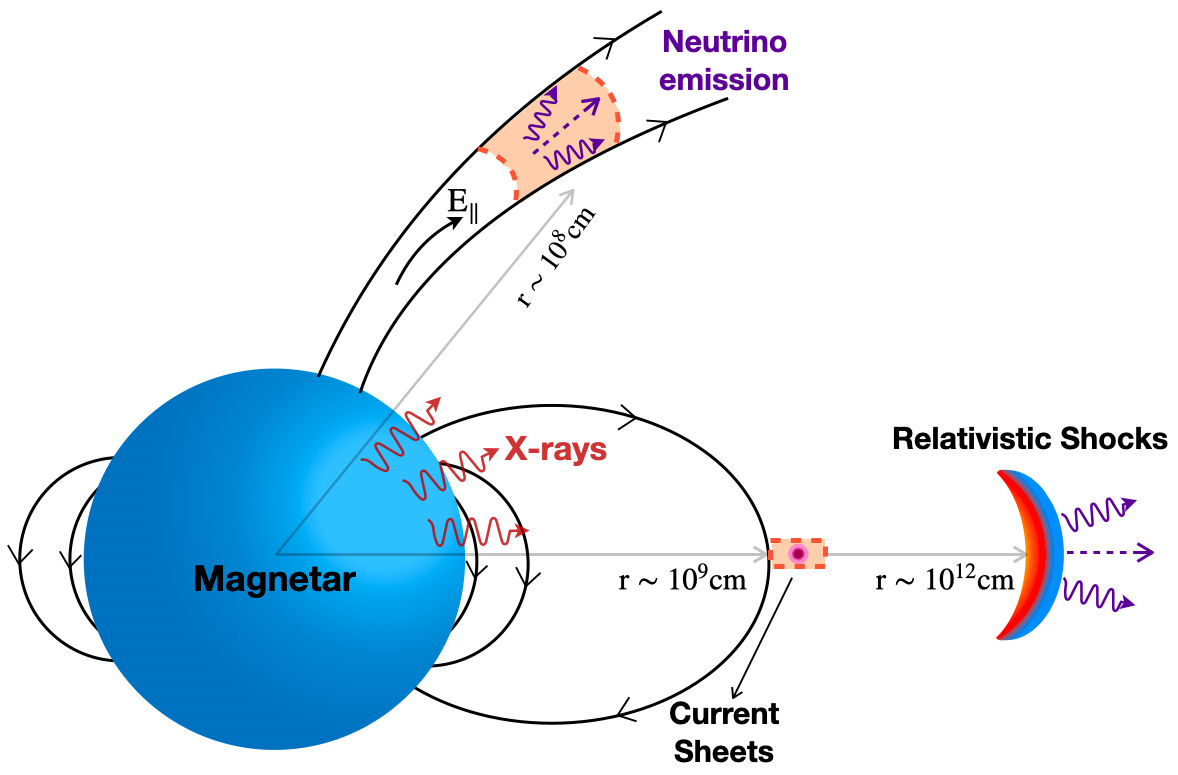}
    \caption{Schematic picture of neutrino production within the magnetosphere, current sheets and relativistic shocks of a flaring magnetar. The purple wiggles or dot (in the case of current sheets) denote the direction of neutrino emission.}
    \label{fig:cartoon}
\end{figure}

For magnetar flares that power X-ray bursts and FRBs, it is likely that some baryons are stripped off the neutron star surface and get accelerated from the magnetar. For simplicity, we assume a hydrogen atmosphere so that protons are stripped and accelerated. In the following, we discuss several mechanisms that may potentially accelerate protons to the desired energies to produce neutrinos.
Figure \ref{fig:cartoon} illustrates the three locations where protons may be accelerated and interact with X-ray photons to produce neutrinos via photomeson interactions.

\subsection{Proton acceleration within the magnetar magnetoshere}

A co-rotating force-free magnetar magnetosphere would carry the Goldreich-Julian charge density \citep{Goldreich&Julian1969}
\begin{equation}
\rho_{\rm GJ}=-\frac{\bm{\Omega}\cdot\textbf{B}}{2\pi{c}}\cdot\frac{1}{[1-(\frac{\Omega{r}}{c})^2\sin\theta^2]}, 
\end{equation}
where $r$ is the radius from the center of the neutron star and $\theta$ is the angle between the rotation axis and the local magnetic field.
When there is a deficit of charge density with respect to $\rho_{\rm GJ}$, a ``gap'' is formed in which a parallel electric field $E_\parallel$ is developed. There are two ways of developing these gaps: One is through binding of surface charges or through a space-charge-limited flow \citep[e.g.][]{Ruderman&Sutherland75,Arons&Scharlemann1979,Muslimov&Harding04}; the other is through propagating Alfv\'en waves to a large altitude from the surface \citep{Kumar&Bosnjak20,Lu20}.

For the FRB model invoking coherent curvature radiation by bunches \citep[e.g.][]{Kumar17,YangZhang18,Lu20}, one requirement is that there exists an $E_\parallel$ to continuously add energy to the emitting bunches to avoid rapid cooling of the bunches \citep{Kumar17}. For this model to work, charged bunches are likely in the radiation reaction limited regime. 

Consider a charged bunch that is balanced by electric field acceleration and curvature radiation cooling:
\begin{equation}
N_e eE_{\parallel}=N_e^2\frac{2e^2\gamma^4}{3\rho^2},
\end{equation}
where $\rho$ is the curvature radius of the magnetic field line and $N_e$ is the number of positrons in a bunch (in order to accelerate protons, $E_\parallel$ needs to have the sign to accelerate positrons). The frequency of curvature radiation is related to Lorentz factor of the bunch and curvature radius of the field line, i.e.
\begin{equation}
\nu=\frac{3c\gamma^3}{4\pi\rho}.
\end{equation}

In order to produce 1-GHz radio waves at the curvature radius of $\rho\simeq10^8 \ \rm{cm}$, one requires $\gamma\simeq241(\rho_8\nu_9)^{1/3}$. Thus the parallel electric field could be expressed as
\begin{equation}
E_{\parallel}\sim(1.1\times10^{4} \ {\rm{esu}}) \ N_{e,20} \rho_8^{-2/3}\nu_9^{4/3}.
\end{equation}

The net charge number density in the magnetosphere may be written as
\begin{equation}
n=\zeta n_{\rm{GJ}}=\frac{\zeta B_{\star}\Omega}{2\pi qc}\left(\frac{r}{R_{\star}}\right)^{-3}\simeq7\times10^{9} \ \zeta_{2}B_{\star,15}R_{\star,6}^3P^{-1}\hat{r}_{2}^{-3},
\end{equation}
where $\zeta$ is the net charge factor normalized to the GJ number density,  $\hat{r}={r}/{R_{\star}}$, $R_{\star}=(10^6 \ \rm cm) R_{\star,6}$ is the radius of magnetar, and $B_{\star}=(10^{15} \ {\rm G}) B_{\star,15}$ is the surface magnetic field of the magnetar. The cross section of a bunch of relativistic particles may be estimated as \citep[e.g.][]{Kumar17} $A=\pi(\gamma\lambda)^2=7.6\times10^8 \ {\rm cm}^2 \ \nu_9^{-2}$, where $\lambda = 30 \ {\rm cm} \ \nu_9^{-1}$ is the wavelength of the FRB emission. The longitudinal size of the bunch is defined by the wavelength of the waves, i.e. $l_{\parallel}\simeq\lambda$.
Thus, the net number of emitting charges in one bunch can be estimated as
\begin{equation}
N_e=n A l_{\parallel}\simeq10^{20} \ \zeta_{2}B_{\star,15}P^{-1}\hat{r}_{2}^{-3}\nu_9^{-3}.
\end{equation}

For an FRB emission site at $\hat r \sim 100$, the characteristic distance for $E_\parallel$ is of the same order, so that the electric potential across the distance may be estimated as $U \sim E_{\parallel} r \sim 1.1\times10^{12} \ N_{e,20}\hat{r}_2\rho_8^{-2/3}\nu_9^{4/3}$. The maximum proton energy gained from this potential can be estimated as
\begin{equation}
\epsilon_{\rm p,max}^{\rm mag} \sim eU \sim
(320 \ {\rm TeV}) \ N_{e,20}\hat{r}_2\rho_8^{-2/3}\nu_9^{4/3}.
\label{eq:epsmax1}
\end{equation}

This energy satisfies the condition for photomeson interaction with 10-200 keV X-rays.

\subsection{Proton acceleration in the current sheet region}
Another proton acceleration site from a magnetar magnetosphere is the current sheet region just outside of the light cylinder, where magnetic field lines with opposite orientations meet and reconnect \citep{Lyubarsky20}. The radius of light cylinder is $R_{\rm LC}=cP/2\pi\simeq4.8\times10^9 \ {\rm cm} \ P$, which corresponds to $\hat r \sim 10^4$. The magnetic field strength at $R_{\rm LC}$ can be estimated as
\begin{equation}
B_{\rm LC}=B_{\star}\left(\frac{R_{\rm LC}}{R_\star}\right)^{-3} \simeq (9.2 \times 10^3 \ {\rm G}) B_{\star,15} P^{-3} R_{\star,6}^3. 
\end{equation}

Let us consider a first-order Fermi acceleration mechanism in the current sheet region. The acceleration timescale in the co-moving frame for proton may be estimated as
\begin{equation}
t_{p,\rm acc} \sim \frac{2\pi\epsilon_p}{eBc}\simeq(4.4\times10^{-5}\epsilon_p \ {\rm s}) \  B_{4}^{-1}.
\end{equation}

The cooling timescale of synchorotron and curvature radiation can be estimated as
\begin{equation}
t_{\rm{cur},p}=\frac{3m_p^4c^7\rho^2}{2e^2\epsilon_p^{3}}\simeq(1.1\times10^{15}\epsilon_p^{-3} \ \rm s) \ \rho_{9}^2,
\end{equation}
\begin{equation}
t_{\rm{syn},p}=\frac{6\pi m_p^4c^3}{\sigma_{{\rm{T}},e}m_e^2 B^{2}\epsilon_p}\simeq(7.2\times10^{7}\epsilon_p^{-1} \ {\rm s}) \ B_4^{-2}.
\end{equation}

In the range of proton energy that is of interest, both radiation cooling timescales are too long compared to the dynamic timescale which is the variability timescale of FRB: $t_{\rm dyn} \sim 1$ ms. One can therefore use the comoving-frame $t'_{\rm p,acc} = t'_{\rm dyn}$ to define the maximum proton energy in the comoving frame. Since both $\epsilon$ and $B$ are smaller by $\Gamma$ in the comoving frame, one has $t'_{\rm p,acc} = t_{\rm p,acc}$. Noticing that $t'_{\rm dyn} = \Gamma t_{\rm dyn} = 0.1 \  {\rm s} \ \Gamma_2 \Delta t_{-3}$, where $\Delta t= (1 \ {\rm ms}) \ \Delta t_{-3}$ is the duration of the FRB, one finally obtains the maximum proton energy in the lab frame: 
\begin{equation}
\epsilon_{\rm p,max}^{\rm cs}(1)\simeq(1.4\times10^5 \ {\rm TeV}) \ \Gamma_2^2B_{\star,15} P^{-3} R_{\star,6}^3\Delta{t}_{-3},
\label{eq:epsmax2}
\end{equation}

Another way of estimating the maximum proton energy in the current sheet region is the one obtained by $\rm{E}_{\parallel}$ in the magnetic reconnection region. According to the numerical simulations of \cite{Guepin20}, the maximum proton energy can reach
\begin{equation}
\epsilon_{\rm p,max}^{\rm cs}(2)\simeq (4.7\times10^6 \ {\rm TeV}) \ B_{\star,15}P^{-1}R_{\star,6}^2,
\end{equation}
where $P$ is the period of the magnetar. This value is larger than $\epsilon_{\rm p,max}^{\rm cs}(1)$. Since $E_\parallel$ in the reconnection region may be screened when dispatched reconnection (magnetic islands) is considered, and since first-order Fermi could occur in reconnection region as well, we will adopt $\epsilon_{\rm p,max}^{\rm cs}(1)$ to perform the current sheet calculations in the rest of the paper.

\subsection{Proton acceleration in relativistic shocks}
Low amplitude Alfv\'en waves excited by magnetar starquakes \citep{Blaes89,Thompson&Duncan1996} can propagate across the magnetosphere to form relativistic pancake-shape ejecta outside the magnetopshere \citep{Yuan20}. Collisions between the ejecta with the magnetar wind \citep{Metzger19} or among the ejecta themselves \citep{Beloborodov20} would drive magnetized relativistic shocks and produce FRB emission via the synchrotron maser mechanism \citep{Sironi21}.  
Since the observed FRB duration is $\Delta t=1 \ \rm{ms}$, the radius of the relativistic shock can be estimated as
\begin{equation}
r_c=2\Gamma^2\Delta tc=(6\times10^{11} \ {\rm{cm}}) \ \Gamma_2^2\Delta t_{-3}.
\end{equation}

Protons can be accelerated in relativistic shocks. The B-field strength can be estimated at the  shock radius $r_c$:
\begin{equation}
B=B_{\rm LC}\left(\frac{r_c}{R_{\rm LC}}\right)^{-1}\simeq 73.6 \ {\rm G} \ B_{\star,15}R_{\star,6}^3\Gamma_2^{-2}\Delta t_{-3}^{-1}P^{-4}.
\end{equation}

The acceleration timescale for protons may be estimated as
\begin{equation}
t_{p,\rm acc} \sim \frac{2\pi\epsilon_p}{eBc}\simeq(5.9\times10^{-3}\epsilon_p \ {\rm s}) \  B_{\star,15}^{-1}R_{\star,6}^{-3}\Gamma_2^{2}\Delta t_{-3}P^{4}.
\end{equation}
Again equating the acceleration and dynamical timescale, one can obtain the maximum proton energy in the lab frame
\begin{equation}
\epsilon_{\rm p,max}^{\rm shock}\sim1.1\times10^3 \ {\rm TeV} \ B_{\star,15}R_{\star,6}^3\Delta t_{-3}^{-1}P^{-4}. 
\label{eq:epsmax3}
\end{equation}

This energy satisfies the condition for photomeson interaction with 10-200 keV X-rays.

\section{Neutrino Emission power}\label{sec:power}
Before examining the neutrino emission spectra in detail, it is informative to estimate the neutrino emission power in three versions of the models. We consider the specific example of FRB200428. Based on observational results of Insight-HXMT \citep{CKLi20}, the unabsorbed X-ray fluence is given by $7.14\times10^{-7} \ \rm{erg} \ \rm{cm}^{-2}$ in the range of 1-250 $\rm{keV}$, and the unabsorbed X-ray flux fitted by a cutoff power law spectral model is given by $8.08\times10^{-7} \ \rm{erg} \ \rm{cm}^{-2} \ s^{-1}$ in the 20-200 keV range. 
Assuming a distance $D\simeq10 \ \rm{kpc}$ from the source, the X-ray luminosity of the FRB-associated burst can be estimated as
\begin{equation}
L_{\rm X}=4\pi D^2 F\simeq(10^{40} \  {\rm{erg} \ \rm{s}^{-1}}) \ D_{\rm 10 kpc}^2 F_{-6}.
\end{equation}

The X-ray photon energy density for three regions (hereafter for Eqs.(\ref{eq:Uph})-(\ref{eq:spect}), from top to bottom: magnetosphere, current sheet, and relativistic shocks, respectively) can be estimated as
\begin{equation}
U_{\rm{ph}}=\frac{L_{\rm X}}{4\pi r^2c}\simeq\left\{
\begin{aligned}
&(2.7\times10^{12} \ {\rm{erg} \ \rm{cm}^{-3}}) \ r_8^{-2}L_{\rm X,40} \\
&(1.2\times10^{9} \ {\rm{erg} \ \rm{cm}^{-3}}) \ P^{-2}L_{\rm X,40} \\
&(2.9\times10^5 \ {\rm{erg} \ \rm{cm}^{-3}}) \ \Delta{t}_{-3}^{-2}\Gamma_2^{-4}L_{\rm X,40}.
\end{aligned}
\right.
\label{eq:Uph}
\end{equation}

X-ray photons will interact with protons to produce neutrinos via photomeson interaction in the three possible acceleration regions. The proton emission power may be estimated as
\begin{equation}
\begin{aligned}
P_{p}&=\frac{4}{3}\sigma_{p\gamma}c\gamma_p^2U_{\rm{ph}}\\
&\simeq\left\{
\begin{aligned}
&(5.3\times10^1 \ {\rm{erg} \ {s}^{-1}}) \ r_8^{-2}\gamma_{p,3}^2L_{\rm X,40}f_{g,0}^2 \\
&(2.3\times10^{-2} \ {\rm{erg} \ {s}^{-1}}) \ P^{-2}\gamma_{p,3}^2L_{\rm X,40}f_{g,0}^2 \\
&(5.9\times10^{-6} \ {\rm{erg} \ {s}^{-1}}) \ \Delta{t}_{-3}^{-2}\Gamma_2^{-4}\gamma_{p,3}^2L_{\rm X,40}f_{g,0}^2.
\end{aligned}
\right.
\end{aligned}
\end{equation}

The photomeson interaction cross section is $\sigma_{p\gamma}\simeq5\times10^{-28} \ \rm{cm}^2$ at the $\Delta$-resonance. We take $\eta_{p\rightarrow\pi}\simeq0.2$ to represent the average fraction of the energy transferred from proton to pion and we multiply $\frac{1}{4}$ to get energy distribution fraction to muon neutrinos \citep{Halzen&Hooper02}: $P_{\nu}\simeq\frac{1}{4}\eta_{p\rightarrow\pi}\frac{4}{3}\sigma_{p\gamma}c\gamma_p^2U_{\rm ph}=0.05P_{p}$. Considering that not all protons 
interact with photons in the observed spectral band, we introduce the correction factor $f_c=1/\ln (\gamma_{\rm p,max}/\gamma_{\rm p,min})$ for each of the three models \citep{Lizhuo2012,Zhang&Kumar2013}. Considering the maximum proton energies in three scenarios (Eqs.(\ref{eq:epsmax1}), (\ref{eq:epsmax2}), and (\ref{eq:epsmax3})) and taking $\gamma_{\rm p,min} \sim 1$ GeV, we get $f_{c}\simeq0.19, 0.09, 0.15$, for the magnetosphere, current sheet and relativistic shock, respectively. The neutrino emission power can be then expressed as
\begin{equation}
P_{\nu}=0.05P_{p}f_c\simeq\left\{
\begin{aligned}
&(0.50 \ {\rm{erg} \ {s}^{-1}}) \ r_8^{-2}\gamma_{p,3}^2L_{\rm X,40}f_{g,0}^2 \\
&(1.05\times10^{-4} \ {\rm{erg} \ {s}^{-1}}) \ P^{-2}\gamma_{p,3}^2L_{\rm X,40}f_{g,0}^2 \\
&(4.42\times10^{-8} \ {\rm{erg} \ {s}^{-1}}) \ \Delta{t}_{-3}^{-2}\Gamma_2^{-4}\gamma_{p,3}^2L_{\rm X,40}f_{g,0}^2.
\end{aligned}
\right.
\end{equation}

According to the observed X-ray burst luminosity and duration $t\approx1.2s$, the total X-ray energy could be estimated as
\begin{equation}
E_{\gamma}=L_{\rm X}t\simeq(1.2\times10^{40} \ {\rm{erg}}) \ D_{10\rm kpc}^2 F_{-6}.
\end{equation}

One may estimate the total isotropic energy of the flare as $E_{\rm{flare}}\simeq10^{41} \ {\rm{erg}} f_{\rm X,-1}^{-1}$ by introducing a radiative efficiency of X-ray emission: $f_{\rm X}=E_{\rm X}/E_{\rm{flare}}=10^{-1} f_{\rm X,-1}$.
The total number of protons can be expressed as
\begin{equation}
N_p=\frac{\epsilon_{\rm acc}E_{\rm{flare}}}{\gamma_p m_p c^2}\simeq6.6\times10^{40} \ \epsilon_{\rm acc}\gamma_{p,3}^{-1}E_{\rm{flare},41},
\end{equation}
where $\epsilon_{\rm acc}$ is the proton acceleration efficiency, which we take as $\sim 1$ in the rest of calculations. The total neutrino luminosities in the three regions can be estimated as:
\begin{equation}
\begin{aligned}
L_{\nu}&=N_pP_{\nu}\\
&\simeq\left\{
\begin{aligned}
&(3.34\times10^{40} \ {\rm{erg} \  s^{-1}}) \ \epsilon_{\rm acc}r_8^{-2}\gamma_{p,3}^2L_{\rm X,40}E_{\rm{flare},41}f_{g,0}^2 \\
&(6.96\times10^{36} \ {\rm{erg} \  s^{-1}}) \ \epsilon_{\rm acc}P^{-2}\gamma_{p,3}^2L_{\rm X,40}E_{\rm{flare},41}f_{g,0}^2 \\
&(2.94\times10^{33} \ {\rm{erg} \  s^{-1}}) \ \epsilon_{\rm acc}\Delta{t}_{-3}^{-2}\Gamma_2^{-4}\gamma_{p,3}^2L_{\rm X,40}E_{\rm{flare},41}f_{g,0}^2.
\end{aligned}
\right.
\end{aligned}
\end{equation}

We assume that the duration of neutrino production is the same as that of X-ray emission: $t\simeq1.2 \ {\rm s}$. The total energy of the radiated neutrinos can be therefore estimated as
\begin{equation}
E_{\nu}=L_{\nu}t\simeq\left\{
\begin{aligned}
&(4.02\times10^{40} \ {\rm{erg}}) \ \epsilon_{\rm acc}r_8^{-2}\gamma_{p,3}^2L_{\rm X,40}E_{\rm{flare},41}f_{g,0}^2 \\
&(8.35\times10^{36} \ {\rm{erg}}) \ \epsilon_{\rm acc}P^{-2}\gamma_{p,3}^2L_{\rm X,40}E_{\rm{flare},41}f_{g,0}^2 \\
&(3.52\times10^{33} \ {\rm{erg}}) \ \epsilon_{\rm acc}\Delta{t}_{-3}^{-2}\Gamma_2^{-4}\gamma_{p,3}^2L_{\rm X,40}E_{\rm{flare},41}f_{g,0}^2.
\end{aligned}
\right.
\end{equation}

At distance $D = 10$ kpc, the neutrino energy fluence can be estimated as
\begin{equation}
\begin{aligned}
\epsilon_{\nu}^2\phi_{\nu}&=\frac{E_{\nu}}{4\pi D^2}\\
&\simeq\left\{
\begin{aligned}
&(3.35\times10^{-6} \ {\rm erg \ {cm}^{-2} \ }, {\rm or} \ 2.1\times10^{-3} \ {\rm GeV 
\ cm^{-2}} ) \  \\
&~~~~~ \times\epsilon_{\rm acc}r_8^{-2}\gamma_{p,3}^2L_{\rm X,40} E_{\rm{flare},41}f_{g,0}^2 D_{\rm 10 kpc}^{-2} \\
&(6.96\times10^{-10} \ {\rm erg \ cm^{-2}}, {\rm or} \ 4.3\times10^{-7} \ {\rm GeV \ cm^{-2}}) \  \\
&~~~~~ \times\epsilon_{\rm acc}P^{-2}\gamma_{p,3}^2L_{\rm X,40} E_{\rm{flare},41}f_{g,0}^2 D_{\rm 10 kpc}^{-2} \\
&(2.94\times10^{-13} \ {\rm erg \ {cm}^{-2}}, \ {\rm or} \  1.8\times10^{-10} \  {\rm GeV \ cm^{-2}}) \  \\
&~~~~~ \times\epsilon_{\rm acc}\Delta{t}_{-3}^{-2}\Gamma_2^{-4}\gamma_{p,3}^2L_{\rm X,40} E_{\rm{flare},41}f_{g,0}^2  D_{\rm 10 kpc}^{-2}.
\end{aligned}
\right.
\end{aligned}
\label{eq:spect}
\end{equation}

The order of magnitude estimates presented here are consistent with the more detailed spectral calculations presented in Section \ref{sec:spect} below.

\section{Neutrino spectra}\label{sec:spect}

In this section, we calculate the predicted neutrino spectra in detail. To do this, we first investigate various cooling timescales of protons in the three emission sites, which are useful to calculate the suppression factors for neutrino emission. 

\subsection{Proton cooling timescales}

As discussed in Section \ref{mechanism},  protons can attain a large Lorentz factor in three acceleration scenarios. During the process of neutrino production, high energy protons also lose energy via two radiative cooling (curvature and synchrotron radiation) processes besides the photomeson hadronic cooling. High energy neutrinos can also be produced via $pp$ interactions, but here we ignore the effect because of the low density environment for the three scenarios. we also ignore the Bethe–Heitler process whose effect is negligibly small. We consider the final proton energy loss timescale being a reduced value of the cooling timescales due to curvature radiation, photomeson and dynamic adiabatic expansion:
\begin{equation}
t_p^{-1}=t_{p\gamma}^{-1}+t_{\rm{cur,p}}^{-1}+t_{\rm{syn,p}}^{-1}+t_{\rm{dyn}}^{-1}.
\end{equation}

Since in the magnetospheric model and probably the current sheet model there is no well-defined bulk motion rest frame for proton emission, we calculate all the timescales in the laboratory frame.

The total curvature radiation power of a single charged particle is given by \citep{Jackson1998}: $P_{\rm{cur}}={2e^2\gamma^4c}/{3\rho^2}$. This does not depend on the mass of the particle, so protons are equally efficient emitters as electrons for curvature radiation. 
The cooling time can be estimated via
$\gamma m_p c^2=P_{\rm{cur}}t_{\rm{cur,p}}$, which gives
\begin{equation}
t_{\rm{cur},p}=\frac{3m_p^4c^7\rho^2}{2e^2\epsilon_p^{3}}\simeq\left\{
\begin{array}{ll}
(3.3\times10^{12} \ {\rm s}) \ \epsilon_{p,\rm TeV}^{-3}\rho_{8}^2, & ~~{\rm magnetosphere},\\
(3.3\times10^{14} \ {\rm s}) \ \epsilon_{p,\rm TeV}^{-3}\rho_{9}^2, & ~~{\rm current ~sheet}.
\end{array}
\right.
\end{equation}

The synchorotron radiation power is given by $P=\frac{4}{3}\gamma^2\sigma_{\rm T,p}c\beta^2U_B$, where $\sigma_{\rm T,p} = \sigma_{\rm T,e}(m_e/m_p)^2$, and $\sigma_{{\rm{T}},e}$ is the Thomson cross section for electron. Thus the synchorotron radiation power of protons is much smaller than that of electrons. The synchrotron
cooling timescale for protons is given by
\begin{equation}
t_{\rm{syn},p}=\frac{6\pi m_p^4c^3}{\sigma_{{\rm{T}},e}m_e^2B^2\epsilon_p}\simeq\left\{
\begin{array}{ll}
(4.8\times10^{7} \ {\rm{s}}) \ \epsilon_{p,\rm TeV}^{-1}B_4^{-2}, & ~~{\rm current ~sheet}, \\
(4.8\times10^{11} \ {\rm{s}}) \ \epsilon_{p,\rm TeV}^{-1}B_2^{-2}. & ~~{\rm shock},
\end{array}
\right.
\end{equation}

Synchorotron cooling is not considered within the magnetosphere model since protons are likely settled in their ground Landau state rapidly. Curvature cooling is not considered in the shock region where the $B$ field curvature is very small. The lab-frame dynamical timescale for proton radiation can be estimated as
\begin{equation}
t_{\rm{dyn}}=(10^3 \ {\rm s}) \ \gamma_{p,3}^2\Delta{t}_{-3},
\end{equation}
where $\Delta t \sim (1 \ {\rm ms}) \Delta t_{-3}$ is the observed typical duration of an FRB. The factor of $\gamma_p^2$ takes into account the conversion from the receiving timescale to the emission timescale. Finally, the reciprocal of the photomeson cooling timescale can be estimated as \citep{Stecker1968,Murase&Nagataki06}
\begin{equation}
t_{p\gamma}^{-1}=\frac{c}{2\gamma_p^2}\int_{{\overline{\epsilon}}_{p,{\rm th}}}^{\infty}\sigma_{p\gamma}(\overline{\epsilon}_\gamma)\kappa_p(\overline{\epsilon}_\gamma)\overline{\epsilon}_\gamma d\overline{\epsilon}_\gamma\int_{\overline{\epsilon}_\gamma/2\gamma_p}^{\infty}\epsilon_{\gamma}^{-2}\frac{dn_{\gamma}}{d\epsilon_{\gamma}}d\epsilon_\gamma,
\end{equation}
where $\sigma_{p\gamma}$ is the cross section of proton-photon scattering, $\kappa_p$ is the inelasticity of proton, $\gamma_p$ is the Lorentz factor of proton, $\bar{\epsilon}_\gamma$ is the photon energy in the rest frame of the proton, and $dn_{\gamma}/d\epsilon_{\gamma}$ is the photon number density in the radiation frame.

\begin{figure}
	\includegraphics[width=\columnwidth]{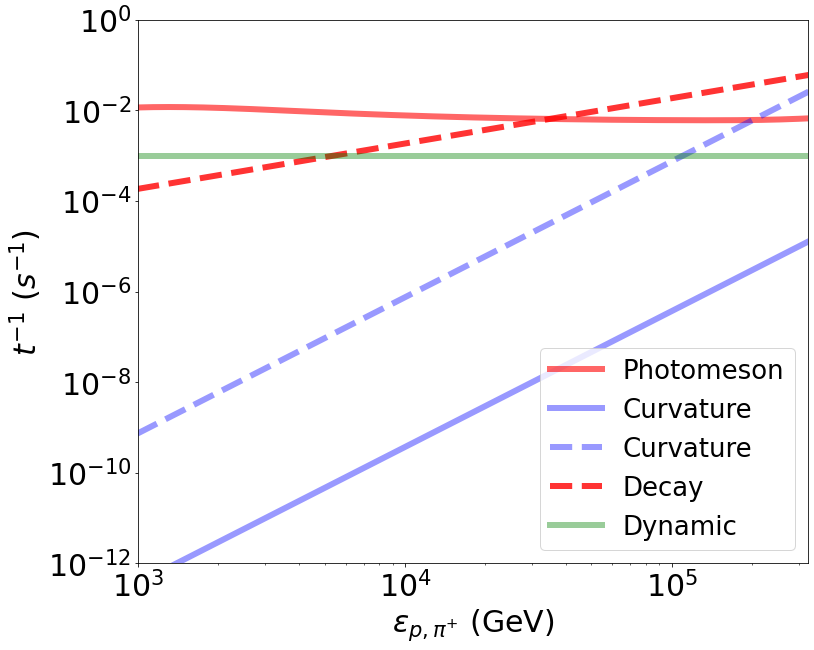}
    \caption{Various $t^{-1}$ terms as a function of energy for protons (solid lines) and pions (dashed lines) in the magnetospheric model. Different processes are plotted in different colors. The parameters are adopted for the Galactic FRB 200428.}
    \label{fig:mag}
\end{figure}

\begin{figure}
	\includegraphics[width=\columnwidth]{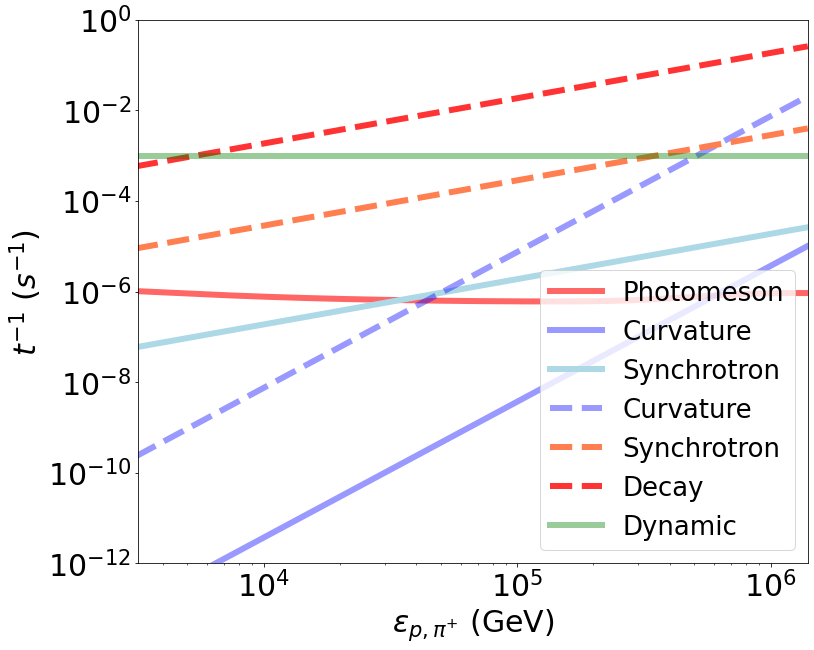}
    \caption{The same as Figure \ref{fig:mag}, but for the current sheet model.}
    \label{fig:sheet}
\end{figure}

\begin{figure}
	\includegraphics[width=\columnwidth]{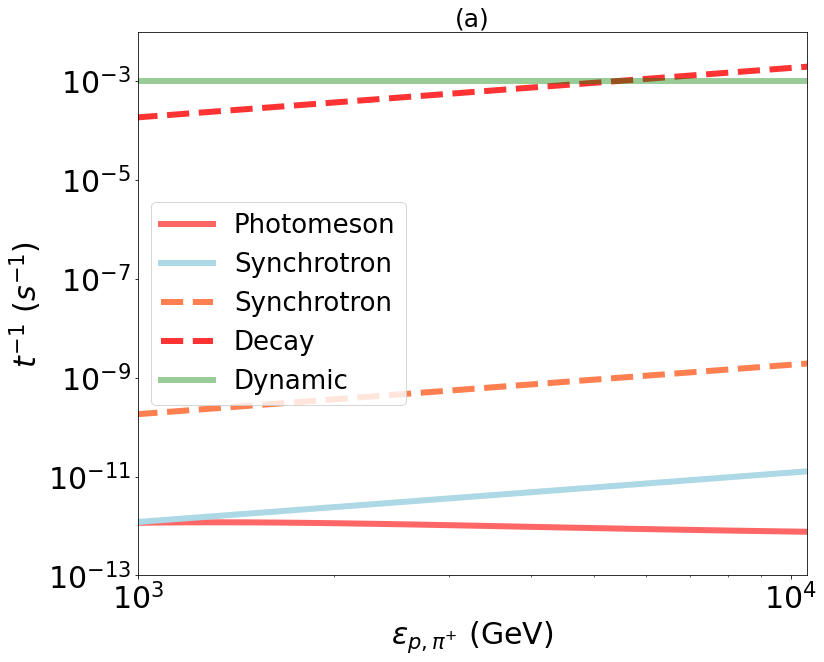}
    \includegraphics[width=\columnwidth]{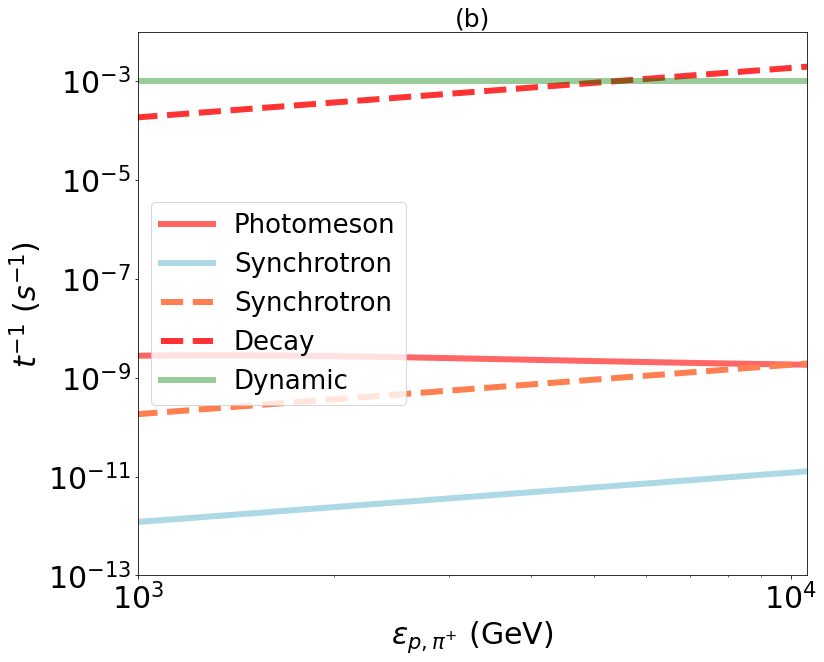}
    \caption{(a) The same as Figure \ref{fig:mag}, but for the relativistic shock model. (b) For $E_{\rm flare}=10^{44} \ {\rm erg \ s^{-1}}$, which is consistent with the parameter adopted by \citep{Metzger20}.}    \label{fig:shock}
\end{figure}

We present various $t^{-1}$ terms for protons for the three acceleration regions using solid lines in Figures \ref{fig:mag}, \ref{fig:sheet} and \ref{fig:shock}, respectively. For the magnetospheric model, $t_{p\gamma}^{-1}$ is larger than $t_{\rm dyn}^{-1}$ and $t_{\rm cuv,p}^{-1}$, suggesting that the $p\gamma$ mechanism is the dominant proton cooling mechanism so that neutrino emission is most efficient. In the current sheet model, $t_{\rm cuv,p}^{-1}$ and $t_{\rm syn,p}^{-1}$ are negligible compared with $t_{\rm dyn}^{-1}$. The latter is greater than $t_{ p\gamma}^{-1}$ so that neutrino emission is suppressed. The same applies to the shock model, with neutrino emission even suppressed. The progressively longer $t_{p\gamma}$ (and hence, the less dominant $t_{p\gamma}^{-1}$) is thanks to the progressively lower target photon density in the $p\gamma$ interaction region, so that neutrino emission power is degraded (see also Section \ref{sec:power}).

Another factor of affecting neutrino spectra is the intermediate pion cooling before pions decay to muons and neutrinos. 
Similar to protons, positive pions also undergo curvature and synchorotron radiation. Their  cooling timescales could be estimated as:
\begin{equation}
t_{\rm{cur},\pi^+}=\frac{3m_{\pi^+}^4c^7\rho^2}{2e^2\epsilon_{\pi^+}^{3}}\simeq\left\{
\begin{array}{ll}
(1.7\times10^{11} \ \rm s) \ \epsilon_{\pi^+,0.2\rm TeV}^{-3}\rho_{8}^2, & ~~{\rm magnetosphere}, \\
(1.7\times10^{13} \ \rm s) \ \epsilon_{\pi^+,0.2\rm TeV}^{-3}\rho_{9}^2, & ~~{\rm current ~ sheet}.
\end{array}
\right.
\end{equation}
\begin{equation}
t_{\rm{syn},\pi^+}=\frac{6\pi m_{\pi^+}^4c^3}{\sigma_{\rm T,e} m_e^2 B^2\epsilon_{\pi^+}}\simeq\left\{
\begin{array}{ll}
(1.1\times10^{5} \ {\rm s}) \ \epsilon_{\pi^+,0.2\rm TeV}^{-1}B_4^{2}, & ~~{\rm current ~sheet}, \\ 
(1.1\times10^9 \ {\rm s}) \ \epsilon_{\pi^+,0.2\rm TeV}^{-1}B_2^{2}, & ~~{\rm shock}.
\end{array}
\right.
\end{equation}
where $\epsilon_{\pi^+}=0.2\epsilon_{p}$ is the pion energy. These should be compared with the decay timescales of pion
\begin{equation}
t_{\rm dec}=\gamma_{\pi^+}\tau_{\pi^+}=2.8\times10^{-8}\gamma_{\pi^+},
\end{equation}
where $\gamma_{\pi^+} = \epsilon_{\pi^+} / m_{\pi^+} c^2$ is the pion Lorentz factor. We present various $t^{-1}$ terms of pions in the three models using the dashed lines in Figs.\ref{fig:mag}, \ref{fig:sheet} and \ref{fig:shock}, respectively.
In all three models, $t_{\rm dec}^{-1}$ is larger than other timescales, suggesting insignificant suppression of neutrino emission during the pion decay phase.

\subsection{Single FRB event neutrino spectrum: the case of FRB 200428}

Photomeson interaction is related to the energy distribution of protons and photons. For simplicity, we assume that the proton distribution obeys a power-law with index $p=2$, i.e.
\begin{equation}
\frac{dN_p}{d\epsilon_p}=\epsilon_p^{-2}.
\end{equation}

We now calculate the neutrino emission from FRB 200428 associated with SGR J1935+2154. Based on the observations, we take the energy distribution of X-ray photons obeying a cutoff power law: $N(\epsilon_{\gamma})d\epsilon_{\gamma}=A\left(\frac{\epsilon_{\gamma}}{100\rm{keV}}\right)^{\alpha}{\rm{exp}}\left(-\frac{\epsilon_{\gamma}}{\epsilon_{\gamma,c}}\right)d\epsilon_{\gamma}$, with $\alpha=-1.56$, cutoff energy $\epsilon_{\gamma,c}=83.89 \ \rm{keV}$, and the normalization factor $A=31.48 \ {\rm cm^{-2} s^{-1}}$ \citep{CKLi20}. The spectrum of photons in the observer frame can be then written as:
\begin{equation}
\frac{dN_{\gamma}}{d\epsilon_{\gamma}}\bigg|_{\rm{obs}}=(31.48  \ {\rm cm^{-2} s^{-1})} \left(\frac{\epsilon_{\gamma}}{100 \ \rm{keV}}\right)^{-1.56}{\rm{exp}}\left(-\frac{\epsilon_{\gamma}}{83.89 \ \rm{keV}}\right).
\end{equation}

When calculating the neutrino spectrum, the photon spectrum in the emission region is relevant. We can convert the observed spectrum to the emission region spectrum by replacing $A_{\rm obs}$ to $A_{\rm src}=A_{\rm{obs}}D^2/(cr^2)$. For the three models, we have  $A_{\rm src}\simeq(1.0\times10^{20} \ {\rm cm^{-3}}) \ D_4^2r_8^{-2}$, $A_{\rm src}\simeq(1.0\times10^{16} \ {\rm cm^{-3}}) \ D_4^2r_{10}^{-2}$, and $A_{\rm src}\simeq(1.0\times10^{10} \ {\rm cm^{-3}}) \ D_4^2r_{13}^{-2}$, respectively. 
In our detailed treatment, we also consider the suppression factors of neutrino emission due to the cooling of protons and pions. For protons, we introduce \citep[e.g.][]{Murase08,Xiao17}:
\begin{equation}
\zeta_{\rm p,sup}=\frac{t^{-1}_{p\gamma}}{t^{-1}_{p\gamma}+t^{-1}_{\rm{dyn}}+t_{\rm{cur,p}}^{-1}+t_{\rm{syn,p}}^{-1}}.
\end{equation}

Similarly, the suppression factor due to pion cooling before decaying is given by:
\begin{equation}
\zeta_{\rm \pi^+, sup}=\frac{t^{-1}_{\rm dec}}{t^{-1}_{\rm{dec}}+t_{\rm{cur,\pi^+}}^{-1}+t_{\rm{syn,\pi^+}}^{-1}}.
\end{equation}

Incorprating the two suppression factors, the neutrino fluence spectrum can be quantified as:
\begin{equation}
\epsilon_{\nu}^2F_{\nu}=\frac{K}{4(1+K)}\times\frac{\epsilon_{\rm acc}E_{\rm{iso}}}{4\pi D_{\rm L}^2{\rm{In}}(\epsilon_{p,\rm max}/\epsilon_{p,\rm min})}\times\zeta_{\rm p,sup}\zeta_{\rm \pi,sup},
\end{equation}
where $K\simeq1$ is adopted for photomeson interactions ($K\simeq2$ is more relevant to proton-proton collisions). The acceleration efficiency is taken as $\epsilon_{\rm{acc}}=1$. The denominator $\ln (\epsilon_{p,\rm max}/\epsilon_{p,\rm min})$ is a normalization factor defined by the minimum and maximum proton energies, which are defined by the threshold and the maximum Lorentz factor of protons in the three scenarios discussed in section \ref{mechanism}.

\begin{figure}
	\includegraphics[width=\columnwidth]{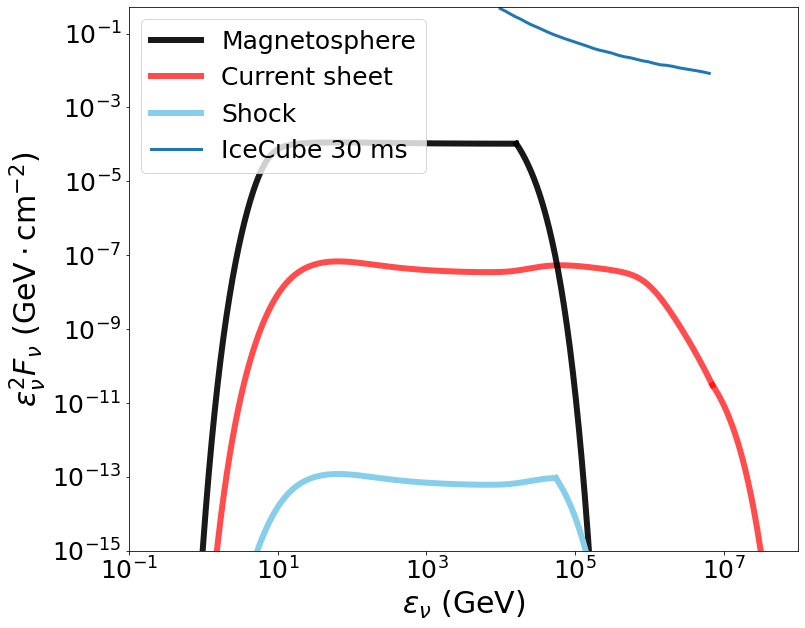}
    \caption{The neutrino spectra of FRB 200428 and its associated X-ray flare from SGR J1935+2154 at $D_L=10$ kpc for the three proton acceleration scenarios. From high to low the three curves are for the magnetosphere (black), current sheet (red) and shock (cyan) models.
    Following parameters are adopted: 
    $E_{\rm{iso}}=10^{41}$ erg, $B_{\star}=2.2\times10^{14}$ G, $\epsilon_{\rm{acc}}=1$. For the magnetosphere, $r=10^8 \ \rm{cm}$, $B=2.2\times10^8 \ \rm G$. For the current sheet region, $r=1.5\times10^{10} \ \rm{cm}$, $B=65.2 \ \rm G$. For the shock model, $r=6\times10^{11} \ \rm{cm}$, $B=1.7 \ \rm G$. The correction factor of the characteristic energy $f_{\xi}=0.3$ is adopted for all three models. IceCube sensitivity curves for observation window of 30 ms \citep{Metzger20} is drawn for comparison. The 1-s sensitivity line (which is more relevant to our calculation) is similar to the 30-ms sensitivity line \citep{Aartsen2020}.}
    \label{fig:FRB200428}
\end{figure}

We present the single event neutrino spectrum from FRB 200428 and its associated X-ray burst from SGR 1935+2154 at $D_{\rm L}=10$ kpc in Figure \ref{fig:FRB200428}. A sharp cutoff in each curve is related to the maximum proton energy for each model (see Section \ref{mechanism}).
We invoke an exponential cutoff function to make the neutrino spectrum smooth at the high energy tail in all three models using the function ${\rm exp}[-(\epsilon_{\nu}-\epsilon_{\nu,\rm max})/f_{\xi}\epsilon_{\nu,\rm max}]$, where $f_{\xi}=0.3$ is the correction factor of the characteristic energy. We also plot the IceCube sensitivity curve for the observation window of 30 ms \citep{Metzger20} for comparison. The 1 s sensitivity curve, which is more relevant to our adopted neutrino emission time, is very close to the 30 ms sensitivity curve \citep{Aartsen2020}. It is clear that even for the most optimistic model for neutrino emission (the magnetosphere model), the predicted spectrum is 4 orders of magnitude too low to be detected by Icecube. The fluxes predicted in the other two models are even lower. In particular, the predicted flux of the shock model is more than 13 orders of magnitude too low. This is qualitatively consistent with \cite{Metzger20}, who suggested that the emission of a Galactic FRB is detectable by IceCube only if the burst is 3 orders of magnitude more energetic and 2 orders of magnitude closer. However, our predicted neutrino flux for the shock model is more than 5 orders of magnitude lower than that of \cite{Metzger20} given the same input parameters. This is because of the small suppression factor $\zeta_{\rm p,sup}\zeta_{\pi,\rm sup}$ we have introduced in the more detailed calculation, In Figure 4b, we present the calculations of various cooling timescales for in the shock model for $E_{\rm flare} = 10^{44} \ {\rm erg \ s^{-1}}$, which is consistent with the parameter adopted by \cite{Metzger20} and is 3 orders of magnitude larger than our adopted fiducial value for FRB 200428. It is clearly shown that $t_{p\gamma}^{-1}$ is more than 5 orders magnitude smaller than $t_{\rm dyn}^{-1}$, so that the suppression factor is $\zeta_{\rm p,sup}\sim2.4\times10^{-6}$.

\section{Diffuse neutrino emission from magnetars}

Finally, it is of interest to calculate the contribution of FRB-emitting magnetars to the observed neutrino background. To do so, we consider all the magnetars in the universe and a range of FRB luminosities (and hence, proton luminosities) to estimate the background emission flux. The diffuse neutrino background radiation from all the sources may be estimated as \citep[e.g.][]{Murase08,Xiao17,Zhu2021}
\begin{equation}
\begin{aligned}
\epsilon_{\nu}^2\Phi&=\frac{c}{4\pi H_0}\int_{0}^{z_{\rm max}}\frac{\rho_{\rm{FRB}}(z)dz}{(1+z)^3\sqrt{\Omega_M(1+z)^3+\Omega_{\Lambda}}}\cdot\frac{K}{4(1+K)}\\
&\cdot\int_{L_{\rm{min}}}^{L_{\rm{max}}}\frac{\Phi(L)\epsilon_{\rm{acc}}E_{\rm{iso}}(L)\zeta_{\rm CRsup}\zeta_{\pi\rm  sup}}{{\rm{In}}(E_{p,{\rm{max}}}/E_{p,{\rm{min}}})}dL,
\end{aligned}
\end{equation}
where $K=1$, $E_{\rm iso} (L)=Lt/\zeta_{\rm eff}$ is the isotropic flare energy. We adopt $t=1.2 \ {\rm s}$ and flare radio efficiency $\zeta_{\rm eff}=10^{-4}$ for all FRBs and the maximum redshift $z_{\rm max}=15$. The luminosity function of FRBs is adopted as a power law as inferred from the data \citep{Luo18,Luo2020,ZhangRC21,DLi21}, i.e. 
\begin{equation}
\frac{dN}{dL}dL\propto\Phi(L)dL\propto L^{-\gamma} dL,
\end{equation}
with $\gamma=1.8$,
$L_{\rm min}=10^{37} \ \rm{erg} \ \rm{s}^{-1}$, $L_{\rm max}=10^{47} \ \rm{erg} \ \rm{s}^{-1}$. Here the rate of FRBs $(\rm{Gpc}^{-3}yr^{-1})$ is assumed to track the star formation history of the universe, which can be parameterized as
\citep{Yuksel08,Sun15}
\begin{equation}
\rho_{\rm{FRB}}(z)=\rho_0 \left[(1+z)^{\alpha\eta}+(\frac{1+z}{B})^{b\eta}+(\frac{1+z}{C})^{c\eta}\right]^{\frac{1}{\eta}},
\end{equation}
where $a=3.4, b=-0.3, c=-3.5, B\approx5000, C\approx9, \eta=-10$, and we take $\rho_0=10^8 \  \rm{Gpc}^{-3}\rm{yr}^{-1}$
\citep{Lu20}. 

\begin{figure}
	\includegraphics[width=\columnwidth]{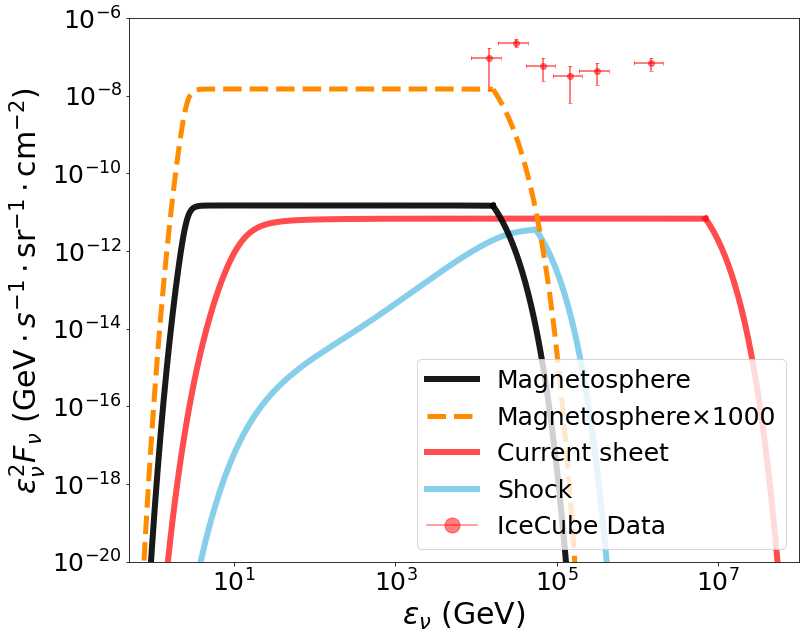}
    \caption{The predicted diffuse background neutrino fluence spectrum for all FRB-emitting magnetars for the three proton acceleration scenarios: magnetosphere (black), current sheet (red), and shock (cyan) models. The orange curve boosts the magnetosphere curve by a factor of 1000, corresponding to the most optimistic case that 1000 times more magnetar X-ray bursts not associated with FRBs also produce neutrinos. The Red data points are the diffuse neutrino backround emission as observed by IceCube. }
    \label{fig:diffuse}
\end{figure}

We show the diffuse background neutrino fluence of all FRB-emitting magnetars in Figure \ref{fig:diffuse}. Similar to Fig.\ref{fig:FRB200428}, the exponential cutoff function is invoked with $f_{\xi}=0.3$ for all three models. One can see that even though the diffuse spectrum of the magnetosphere model is similar to that for single event, those of the current sheet and shock models have different shapes and have closer fluxes to the magnetosphere model. This is because when FRB luminosity function is considered, brighter FRBs would have larger flare energies and also larger X-ray luminosities, so that the suppression factors for the latter two models become larger and closer to unity. The flat spectra in the magentosphere and current sheet models suggest that proton emission efficiency has saturated to the maximum value, while the rising spectrum in the shock model suggests the progressively increasing suppression factors towards unity as one goes to higher energies. 

Nonetheless, even for the most optimistic model invoking proton acceleration in the magnetosphere, the predicted neutrino diffuse emission only reaches  $10^{-11} \ \rm GeV \ s^{-1} \ sr^{-1} \ cm^{-2}$, which is negligibly small compared with the observed cosmic neutrino background. On the other hand, observations \citep{Lin20} showed the majority of magnetar X-ray bursts are not associated with FRBs. It is unclear whether these X-ray flares can accelerate protons and produce neutrinos. If we assume that they can do so, the predicted neutrino background is boosted. The orange dashed line of Fig.\ref{fig:FRB200428} shows the revised diffuse background assuming that there is a factor of 1000 more magnetar X-ray bursts that are not associated with FRBs but still contribute to neutrino background emission. One can see it reaches the level of $10^{-8} \ \rm GeV \ s^{-1} \ sr^{-1} \ cm^{-2}$ in GeV to multi-TeV range.

\section{Conclusions and Discussion}

In this paper, we have proposed a general framework to study neutrino emission from FRB-emitting magnetars and discussed three scenarios of proton acceleration in detail. Our conclusions can be summarised as follows:
\begin{itemize}
    \item The three scenarios invoke three different sites of proton acceleration:  the inner magnetosphere of the magnetar,  the current sheet region just outside the light cylinder of the magnetar, and the relativistic shocks far from the magnetosphere. All three sites allow protons to be accelerated to high enough energy to interact with $10-200$ keV X-ray photons to produce TeV neutrinos. 
    \item The flux of the neutrino emission progressively decreases with increasing emission radius from the magnetar. The main reason is that the target photon number density decreases rapidly with radius, so that neutrino emission optical depth decreases accordingly. 
    \item We calculated the neutrino flux of FRB 200428 and its associated X-ray burst from the Galactic magnetar SGR J1935+2154. We found that even for the most optimal case involving magnetospheric proton acceleration, the predicted flux is still $\sim$ 4 orders of magnitude below the IceCube sensitivity. More energetic Galactic FRBs from closer distances may be detected by IceCube in the future within the magnetospheric scenario. The detection or non-detection of these future events may provide a diagnosis on the emission site of particle acceleration in FRB-emitting magnetars.
    \item We calculated the diffuse neutrino background emission from FRB-emitting magnetars. Even under the most optimistic magnetosphere model, the FRB-emitting magnetars in the universe provide a negligible contribution to the cosmic neutrino background during the bursting phase. However, if one assumes that many more X-ray bursts without FRB associations can also produce neutrinos with the similar mechanism(s), FRB-emitting magnetars can contribute up to $10^{-8} \ \rm GeV \ s^{-1} \ sr^{-1} \ cm^{-2}$ in GeV to multi-TeV range.
\end{itemize}

\section*{Acknowledgements}

We thank Shunke Ai, Ke Fang, Brian Metzger, and Jinping Zhu for helpful discussion and an anonymous referee for suggestions.

\section*{Data Availability}

The code developed to perform the calculation in this paper is available upon request.








\bsp	
\label{lastpage}
\end{document}